\documentclass[amsmath,floatfix]{revtex4}

\usepackage{graphicx}

\newcommand{\be}{\begin{equation}}
\newcommand{\ee}{\end{equation}}
\newcommand{\bea}{\begin{eqnarray}}
\newcommand{\eea}{\end{eqnarray}}
\newcommand{\ba}[1]{\begin{array}{#1}}
\newcommand{\ea}{\end{array}}

\begin{document}

\preprint{MIT-CTP-3745}

\title{Higgs-field Portal into Hidden Sectors} 

\author{Brian Patt}
\email{blpatt@mit.edu}
\author{Frank Wilczek}
\email{wilczek@mit.edu}
\affiliation{Center for Theoretical Physics, Massachusetts Institute 
of Technology, Cambridge, MA 02139, USA}

\date{\today}

\begin{abstract}
The Higgs field mass term, being superrenomalizable, has a unique status within the standard model.   Through the opening it affords, $SU(3) \times SU(2) \times U(1)$ singlet fields can have renormalizable couplings to standard model fields.   We present examples that are neither grotesque nor unnatural.  A possible consequence is to spread the Higgs particle resonance into several weaker ones, or to afford it additional, effectively invisible decay channels.
\end{abstract}


\maketitle

With one exception, all the interactions of the standard model are associated with strictly renormalizable interactions.  In other words, the interactions and kinetic terms are represented by operators of mass dimension 4; or, in still other words, their associated couplings are dimensionless, in units with $\hbar = c =1$.   The Higgs field mass term $\Delta {\cal L} = - \mu^2 \phi^\dagger \phi$, with a coupling of mass dimension 2 and an interaction operator of mass dimension 2, is the exception.   Due to that circumstance, the Higgs field is uniquely open to renormalizable (or superrenormalizable) coupling to $SU(3)\times SU(2) \times U(1)$ singlet fields.  Of course, neither Higgs particles nor $SU(3)\times SU(2) \times U(1)$ singlets have been observed as yet: the former, presumably because they are too heavy for existing accelerators; the latter, because no generating source has been available.   The preceding observation invites us to speculate that their appearance might be contemporaneous.

Several theoretical ideas motivate the concept of  ``hidden'' sectors consisting of $SU(3)\times SU(2) \times U(1)$ singlet fields.   Independent of any model, we can simply note that known standard model fields couple to a variable number of the standard model gauge fields, from all three for the left-handed quark fields to just one for the right-handed electron.  (And in the seesaw mechanism of neutrino mass generation, heavy $SU(3)\times SU(2) \times U(1)$ singlet fermion fields play a crucial role.)  Thus there is no evident reason, if we envisage additional product gauge fields, not to imagine that there are fields which transform under the new but not under the familiar gauge symmetries.  Specific models containing fields that could be construed as a forming a (tiny) hidden sector have been considered for phenomenological purposes \cite{dklm}.  The product $E(8)\times E(8)$ structure, with forms of matter transforming only under one or the other factor, arises in heterotic string theory \cite{ghmr}, and many other string theory constructions also lead to structures of that sort.   

Most discussion of hidden sectors has posited that they are associated with a high mass scale.   That assumption immediately explains why the ``hidden'' sector is in fact hidden, but leaves the challenge -- a form of hierarchy problem -- of understanding why interactions do not pull the mass scale of the visible sector close to that high scale.   But it is equally conceivable that the intrinsic scale of the hidden sector is smaller than, or comparable to, that of the visible sector.   As we shall see immediately below, in that case coupling of the hidden to the visible sector generally occurs only through fields whose masses are naturally pulled up close to the visible scale (what was a bug in the other direction becomes, read this way, a feature).    In this way we find a simple explanation for why a light ``hidden'' sector could in fact have remained hidden to date.   It need not remain so, however, once the Higgs portal opens.   

To distinguish the specific type of hidden sector we discuss here, that is with approximately weak scale or lighter fields coupled to standard model fields only through the Higgs (mass)$^2$ term, we shall henceforth speak of phantom matter.

\bigskip

{\it Simplest Models}: A very simple phantom sector, coupled to ordinary matter only through the Higgs mass term, could implement the attractive idea that fundamental interactions contain no explicit mass scale at all.  Indeed, we can let the phantom sector consist of a confining massless gauge theory that ``totally'' commutes with  the standard model (s) $SU(3)_s \times SU(2)_s \times U(1)_s$, in the sense that its quarks are $SU(3)_s \times SU(2)_s \times U(1)_s$ singlets.   The effective theory of the hidden sector will be a sort of $\sigma$ model, and the phantom (p) $\sigma_p$ field will couple to the standard model in the form ${\cal L}_{\rm link} =  \eta \phi^\dagger_s \phi_s \sigma_p^2$.   Linear terms are forbidden by phantom chiral symmetry.   Spontaneous chiral symmetry breaking with $\langle \sigma_p \rangle = \kappa$ will then generate an effective mass$^2$ $-\eta \kappa^2$ for $\phi_s$, which, assuming $\eta > 0$, could trigger electroweak symmetry breaking.   In this scenario the ratio between the weak scale and the Planck scale would arising from an effect in the phantom sector similar to that which (presumably) works in our strong sector to generate the ratio between the proton mass and the Planck scale \cite{mass}.   An enormous disparity of scales can be required in order for a moderate value of the phantom gauge coupling at the Planck scale to evolve to a large value, and induce chiral symmetry breaking, because running of couplings is logarithmic.   As in that case, no terribly small (unnatural) numerical quantities need be involved.   Of course, one would still have to understand why the intrinsic mass$^2$ vanishes, or is subdominant.   

As another simple and easily comprehended example, consider a doubled version of the standard model \cite{bgh}.  We have the gauge group $SU(3)_{\rm s}\times SU(2)_{\rm s} \times U(1)_{\rm s}\times SU(3)_{\rm p}\times SU(2)_{\rm p}\times U(1)_{\rm p}$, with standard model (s) left-handed quarks in $(3, 2, -\frac{1}{6}, 1,1,0)$, phantom (p) quarks in $(1,1,0, 3, 2, -\frac{1}{6})$, and so forth.   We will not however demand that the numerical values of the phantom sector parameters are equal to ours.   In line with the preceding discussion, the only way that these worlds communicate with one another is through the term ${\cal L}_{\rm link} =  \eta \phi^\dagger_s \phi_s \phi^\dagger_p \phi_p$ linking the two Higgs doublets fields.    (For our purposes we ignore the possibility of cross-coupling between the $U(1)_{\rm s}\times U(1)_{\rm p}$ gauge fields, which is special to this model .  That coupling must be small on phenomenological grounds.   It could be forbidden by imposing $(CP)_p$ symmetry.) 

To assess the significance of the linking term, we must analyze the complete potential
\begin{equation}
V(\phi_s, \phi_p) = \mu_s^2 \phi_s^\dagger \phi_s + \lambda_s  (\phi_s^\dagger \phi_s)^2 + \mu_p^2 \phi_p^\dagger \phi_p + \lambda_p  (\phi_p^\dagger \phi_p)^2 -  \eta \phi^\dagger_s \phi_s \phi^\dagger_p \phi_p
\end{equation}
To insure $SU(2)_s\times U(1)_s$ breaking, we demand that this is minimized at $\langle \phi_s \rangle = v_s \neq 0$.   The algebra for a general analysis gets quite unpleasant, but the main qualitative results are easily described.  There are two distinctly different phenomenologies, depending upon whether $\phi_p$ does or does not acquire a non-zero vacuum expectation value.  

If $\langle \phi_p \rangle = v_p \neq 0$ then as we expand $\phi_s = v_s + h_s, \phi_p = v_p + h_p$ to isolate the fields $h_s, h_p$ that produce quanta of definite mass above the ground state, we find -- ignoring vacuum energy and linear terms, which of course must cancel -- cross-couplings
\begin{equation}
{\cal L}_{\rm cross} =  \eta ( 2 v_s h_s  + h_s^2)( 2 v_p h_p  + h_p^2)
\end{equation}
This includes a quadratic term $\eta 4 v_s v_p h_s h_p$, which induces mixing.  The mass eigenstates $h_1, h_2$ will be linear combinations of $h_s$ and $h_p$.    Since only $h_s$ couples directly to ordinary matter, the two particles of definite mass will share the production rate that one would have calculated for the standard model Higgs particle.    Furthermore, there will be new channels open for decay, through couplings of $h_p$ to the new, $SU(3)_s \times SU(2)_s \times U(1)_s$ singlet phantom particles.  Unless special, arduous effort were invested toward their detection, the phantoms would be perceived only as missing energy and momentum.     In principle, one could imagine observing energy re-deposited downstream as the new matter scatters off ordinary matter through Higgs (i.e., $h_1$ and $h_2$) exchange, by adapting techniques devised for WIMP detection.  

If $\langle \phi_p \rangle = 0$ then there is no quadratic term, and $h_s, h_p$ are the mass eigenstates.  Moreover $h_p$ is one component of a degenerate quartet, all with the same coupling to $h_s$.   The remaining couplings
\begin{equation}
{\cal L}_{\rm cross} =  \eta ( 2 v_s h_s  + h_s^2)h_p^2
\end{equation}
can lead to the invisible decay $h_s \rightarrow h_p h_p$ if $h_p$ is sufficiently light.   If that channel is not kinematically allowed, there still might be invisible decays through virtual $h_p$.   One would also have the possibility to observe missing energy through virtual $h_s^* \rightarrow h_p h_p$, or the bremsstrahlung-like process $h_s^* \rightarrow h_s h_p h_p$.   Overall, the phenomenology in this case is much less striking.

Note that in both cases the $h_p$ fields get contributions of the type $\eta v_s^2 h_p^\dagger h_p$ to their mass$^2$ matrix.   This exhibits the tendency for their masses to be pulled  toward the visible sector scale, that we advertised earlier.

\bigskip

{\it Possible Generalizations and Implications}:

\begin{enumerate}

\item 

In alternative models there could be not only one but several fields from the $SU(3)_s \times SU(2)_s \times U(1)_s$ singlet sector that mix with the conventional Higgs field.   Then the conventional production rate would be shared among several different states, and conventional signals (apart from missing energy-momentum and downstream interactions) would be further diluted by decays into the phantom sector.  These effects would make the physics of the Higgs sector richer, but more challenging to access.  

\item

It is useful to classify possible couplings types of $SU(3)_s \times SU(2)_s \times U(1)_s$ singlet scalars into three categories.   Categories 1 and 2 arise when these scalars are non-singlets under some phantom gauge group.  Such visible singlet, phantom non-singlet fields couple in the format ${\cal L}_{\rm link} =  \eta \phi^\dagger_s \phi_s \phi^\dagger_p \phi_p$ we encountered above, where by $\phi_p^\dagger \phi_p$ we generally mean an invariant sum over a multi-component field.    If $\phi_p$ does not acquire a vacuum expectation value (category 1), the phenomenology will be similar to what was discussed in the penultimate paragraph of the previous section.  If $\phi_p$ does acquire a vacuum expectation value (category 2), the phenomenology will be similar to what was discussed just before that, with mixing.   In general there may be components of $\phi_p$ that are not absorbed in the hidden sector Higgs mechanism.  Those extra components, if any, will couple similarly to the category 1 $h_p$.  Category 3 allows general couplings of the singlet, including the trilinear $\phi_p \phi^\dagger_s \phi_s$.  This could only arise as a primary interaction if $\phi_p$ has no non-trivial gauge symmetry transformation.  

The right-handed neutrino $N$ in the seesaw mechanism of neutrino mass generation is a sort of category 3 fermion field, which couples to the $SU(3)_s \times SU(2)_s \times U(1)_s$ dimension $\frac{5}{2}$ fermionic invariant $\phi_s^\dagger L_s$, where $L_s$ is  a lepton doublet, to make the dimension 4 Lagrangian term $\phi_s^\dagger L_s N$.  This is then integrated out, for a heavy $N$, to give the effective dimension 5 (non-renormalizable) term $(\phi_s^\dagger L_s)^2$.  Category 1 or 2 phantom fermion fields can have no renormalizable couplings to standard model fields, but one could entertain the possibility of generalizing the seesaw mechanism to include hidden sectors, e.g. with dimension 5 terms of the form $\Delta {\cal L} \propto L_p^\dagger \phi_p \phi^\dagger_s L_s$.   Such terms could induce phenomena analogous to what we have mentioned for the Higgs sector, i.e. admixture of phantom -- effectively, ``sterile'' -- components, now in the neutrino sector.

\item

The phenomenology of our first model falls into category 2.  A special feature is that the massless Nambu-Goldstone bosons associated with phantom chiral symmetry breaking will couple directly to the observable Higgs particle(s), and could dominate their decay.  

\item

Cosmology offers both opportunities for, and constraints upon, the ideas discussed here.  Evidently, many additional possibilities for dark matter arise.  The success of cosmological nucleosynthesis calculations puts severe constraints upon the existence of additional degrees of freedom that are in thermal equilibrium with, or have an equal or greater temperature than, the known standard model particles at the time of nucleosynthesis.   Phantom hidden sectors of our sort will tend to fall out of equilibrium with the standard model particles once the Higgs particles have disappeared.  Thus hidden sector matter will not participate in the adiabatic heating associated with later disappearance episodes (e.g., muon-antimuon and electron-positron annihilation).   The phantom sector could therefore be substantially cooler than ordinary matter at the time of nucleosynthesis, but only if the phantom spectrum obeys appropriate constraints.

\item

Adapting these ideas to models with low-energy supersymmetry brings in many additional considerations and choices.  Here we confine ourselves to two simple comments.   First, the mixing phenomena discussed here can only affect the neutral Higgs fields.   Second, since we are adding only $SU(3)_s \times SU(2)_s \times U(1)_s$ singlets, there need be only minimal perturbation of the successful unification of couplings calculation.

\item

In earlier work we discussed the possibility of a hidden sector implementing spontaneous breaking of family symmetry at a high mass scale \cite{flavor}.  We found that in some models, for reasons tied up with supersymmetry, non-derivatively coupled scalar fields from that sector could have masses of order the weak scale.  This is another, more intricate scenario that might lead to an effective phantom sector.

\end{enumerate}

\begin{acknowledgments}
We thank R. Jaffe, I. Stewart and especially D. Tucker-Smith for helpful comments. 
FW acknowledges support from the U. S. Department of Energy contract DE-FG02-05ER41360. 
\end{acknowledgments}

\end{document}